\documentclass[12pt]{article}
\usepackage{amsmath,amssymb}
\usepackage{eucal}

\usepackage[dviwindo]{graphicx}

\newcommand{\bs}{\begin{subequations}}
\newcommand{\es}{\end{subequations}}
\numberwithin{equation}{section}
\newcommand{\ben}{\begin{eqnarray}}
\newcommand{\een}{\end{eqnarray}}
\newcommand{\la}{\label}
\begin{document}

\title{Exact Solutions of Regge-Wheeler Equation and Quasi-Normal Modes of Compact Objects}

\vskip 1.5truecm

\author{Plamen~P.~Fiziev\thanks{Department of Theoretical Physics, University of
Sofia, Boulevard 5 James Bourchier, Sofia 1164, Bulgaria;
E-mail:\,\,\,fiziev@phys.uni-sofia.bg}}

\date{}
\maketitle

\begin{abstract}
The well-known Regge-Wheeler equation describes the axial
perturbations of Schwarzschild metric in the linear approximation.
From a mathematical point of view it presents a particular case of
the confluent Heun equation and can be solved exactly, due to
recent mathematical developments. We present the basic properties
of its general solution. A novel analytical approach and numerical
techniques for study the boundary problems which correspond to
quasi-normal modes of black holes and other simple models of
compact objects are developed.
\end{abstract}

\sloppy

\section{Introduction}

In the linear approximation the axial perturbations of
Schwarzschild metric of Keplerian mass $M$ are described by the
well-known Regge-Wheeler equation \cite{RW}:
\ben  \partial_t^2 \Phi_{s,l} +\left(-\partial^2_{x}
+V_{s,l}\right)\Phi_{s,l} = 0. \la{RW}\een
Here $x=r+r_{\!{}_S}\ln\left(r/r_{\!{}_S}-1\right)$ is the
Regge-Wheeler "tortoise" coordinate, $r_{\!{}_S}=2M(=1)$ is the
Schwarzschild radius, and $\Phi_{s,l}(t,r)$ is the radial function
for perturbations of spin $s$ and angular momentum $l\geq s$. The
corresponding effective potential $V_{s,l}(r)$ reads:
$V_{s,l}(r)=\left(1-{1\over r}\right)\left({{l(l+1)}\over {r^2}}
+{{1-s^2}\over {r^3}} \right)$.  The area radius $r$ in it can be
expressed explicitly as a function of the variable $x$ using
Lambert W-function \cite{W}: $r/r_s=LambertW(e^{x/r_s-1})+1$.

The standard ansatz $\Phi_{s,l}(t,r)=R_{\varepsilon,s,l}(r)
e^{i\varepsilon t}$ brings us to the stationary problem:
\ben
\partial^2_x R_{\varepsilon,s,l}+\left(\varepsilon^2-V_{s,l}\right)
R_{\varepsilon,s,l}=0.
\la{R}\een

A large amount of references on this subject can be found in the
review articles \cite{QNM}. There we meet often the statement that
the exact solution of the problem is not known. Therefore, quite
sophisticated approximate analytical and numerical techniques for
solution of physical problems related to Eqs. (\ref{RW}),
(\ref{R}) were developed in the past. They discovered rich and
important mathematical and physical properties of the
corresponding problems. Some of them need further justification.

\begin{figure}[htbp] \vspace{6.truecm}
\includegraphics{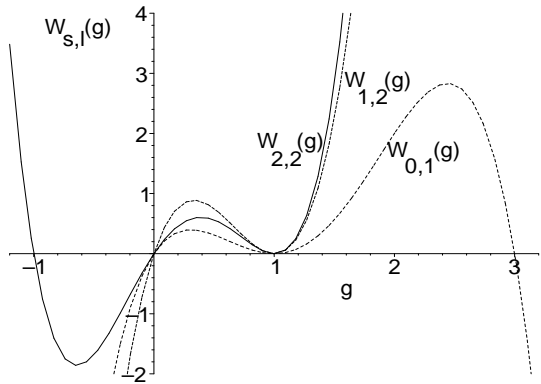} \caption{\hskip 0.2truecm Typical forms of the
potentials $W_{s,l}(g)$ for different $s,l$. The physical domain
is $g\in (0,1)$. The point $g=0$ represents the Schwarzschild
event horizon, the point $g=1$ -- the physical infinity.
    \hskip 1truecm}
    \label{Fig0}
\end{figure}

A hint on the class of new mathematical functions, which solve the
stationary Regge-Wheeler problem (\ref{R}) is the use of the very
time-time component of Schwarzschild metric $g_{tt}=1-1/r=g$ as an
independent variable, instead of the area-radius $r$. This
transforms the Eq. (\ref{R}) into:
\ben
\partial^2_{x}R_{\varepsilon,s,l}+\left(\varepsilon^2-W_{s,l}\right)
R_{\varepsilon,s,l}=0 \la{Pg}\een
with the potential $W_{s,l}(g)=l(l+1)g(1-g)^2+(1-s^2)g(1-g)^3$
which describes anharmonic oscillators of different type,
depending on the values of spin: $s=0$ -- for scalar waves, $s=1$
-- for electromangnetic waves, $s=2$ -- for gravitational waves
(see Fig. \ref{Fig0}).

This simple observation sheds additional light on the physical
content of Eq.(\ref{R}) and yields a better understanding of the
qualitative differences and similarities between the perturbations
of different spin. In particular, one can see in a more clear way
the behavior of the potential, the RW equation and its solutions
at infinite space points when these are placed at finite
distances, using the variable $g$.

Here we utilize the form (\ref{Pg}) of the problem at hand only as
a hint on a new direction of mathematical investigation.  It is
well known that the above polynomial form of the potential
$W_{s,l}(g)$ leads to a solution of geodesic equations in
Schwarzschild metric in elliptic functions. According to the early
article \cite{Manning}, in this case the corresponding
Schr\"odinger like spectral problem can be solved by the method of
continuous fractions. This method was developed for the specific
RW spectral problem by Leaver \cite{QNM} and at present is one of
the basic methods for finding the QNM spectra.

The modern mathematical development shows that many physical
problems, including anharmonic oscillator ones, are solved exactly
by Heun's functions \cite{Heun}. Thus the form of the
Eq.(\ref{Pg}) suggests to describe the exact solution of
Regge-Wheeler equation using Heun's functions. We shall show that
the Eqs. (\ref{R}), (\ref{Pg}) can be transformed to a particular
case of confluent Heun equation.

The confluent Heun equation of general form is well known and
studied in some detail in recent mathematical literature
\cite{Heun}. At present its solutions are more or less
well-described special functions, already included in modern
computer packages. In particular, the Heun's functions are
included in the last version of the widespread computer package
Maple 10, which has been used in all original numerical
calculations in our article. These functions present a nontrivial
generalization of familiar hypergeometric functions and have much
richer and complicated properties, since the Heun equation has one
more singular point, than the hypergeometric equation.

In the present article, we give the exact solution of the
Regge-Wheeler equation in terms of Heun's functions and apply them
for study of different boundary problems in the relativistic
theory of gravity.

\section{Solutions of Regge-Wheeler Equation}

\subsection{Local Solutions of Confluent Heun Equation}

It is convenient to use the standard variable $r$ for reaching the
standard canonical form of the Heun equation of our problem. The
explicit form:
\ben {{d^2R_{\varepsilon,s,l}}\over{dr^2}}+
{1\over{r(r-1)}}{{dR_{\varepsilon,s,l}}\over{dr}}+
{{\varepsilon^2r^4-l(l+1)r^2+\left(l(l+1)+s^2-1\right)r+1-s^2}\over
{r^2(r-1)^2}}R_{\varepsilon,s,l}=0\,\,\, \la{P}\een
of Eq. (\ref{R}) shows that it has three singular points: the origin
$r=0$, the horizon $r=1$ and the infinite point $r=\infty$
\cite{Heun}. The first two are regular singular points and may be
treated on equal terms. The last one is an irregular singular point,
obtained as a result of confluence of two regular singular points in
the general Heun equation. Hence, we are to consider a specific
representative of the class of confluent Heun equations. This
circumstance is the source of complicated analytical properties of
solutions of the Regge-Wheeler equation, as we shall see below.

For transformation of Eq. (\ref{P}) to a canonical form of the
Heun equation \cite{Heun} :
\ben {{d^2H}\over{dr^2}}+\left(\alpha+{{\beta+1}\over{r}}+
{{\gamma+1}\over{r-1}}\right){{dH}\over{dr}}+\hskip 5.3truecm\nonumber\\
{1\over{r(r-1)}} \left( \Big(\delta+{1\over
2}\alpha(\beta+\gamma+2)\Big) r+\eta+{\beta\over 2}+{1\over
2}(\gamma-\alpha)(\beta+1) \right)H=0\la{H}\een
one may use the ansatz
$R_{\varepsilon,s,l}(r)=r^{s+1}(r-1)^{i\varepsilon} e^{i\varepsilon
r}H(r)$. It yields the following specific parameters:
$\alpha=2i\varepsilon, \beta=2s, \gamma=2i\varepsilon,
\delta=2\varepsilon^2, \eta=s^2-l(l+1)$.

There exist several substitutions of that kind. All they transform
Eq. (\ref{P}) to the same form (\ref{H}), but with different
coefficients, for example, with all possible combinations of the
signs of the coefficients $\alpha$, $\beta$ and $\gamma$. The
corresponding solutions of equations (\ref{H}), with different
coefficients, are related by simple transformations, described in
literature \cite{Heun}. It is useful to know all sets of such
transformations and corresponding parameters. In some cases this
permits us to construct two independent solutions of stationary
Regge-Wheeler equation (\ref{R}), using the basic standard
canonical solution of confluent Heun equation
(\ref{H})\footnote{In present article we apply the notations for
Heun's functions, which are used in Maple 10.}:
\ben HeunC\left(\alpha,\beta,\gamma,\delta,\eta,r\right),\la{Scan}
\een
which is defined by the convergent Taylor series expansion around
the origin $r=0$ and normalization:
\begin{subequations}\label{norm:ab}
\ben
HeunC\left(\alpha,\beta,\gamma,\delta,\eta,0\right)=1,\hskip 5truecm\la{norm:a}\\
{d\over{dr}}HeunC\left(\alpha,\beta,\gamma,\delta,\eta,0\right)=
{1\over 2}
{{\beta(-\alpha+\gamma+1)-\alpha+\gamma+2\eta}\over{\beta+1}}.\la{norm:b}\een
\end{subequations}

The Taylor series expansion of the standard canonical solution
(\ref{Scan}) with respect to the independent variable $r$ is
obtained using the known three-terms recurrence relation
\cite{Heun} and initial conditions (\ref{norm:ab}).

It is not excluded that some specific sets of parameters $\alpha,
\beta, \gamma, \delta, \eta$ may not yield well defined canonical
solutions (\ref{Scan}) of the confluent Heun equation (\ref{H}).
This indeed happens in our problem, since in it $s\in\mathbb{Z}$
and $s\geq 0$. As a result, for example, solutions
$HeunC\left(\alpha,\beta,\gamma,\delta,\eta,r\right)$ with
$\beta=-2s$ are not well defined, because the denominator of the
coefficients in Taylor series expansion includes products of form
$\beta(\beta+1)...(\beta+N), N\in \mathbb{Z}^+$.

One can go outside the domain of convergence of the Taylor series
expansion -- the open unit disk in $\mathbb{C}_r$, centered at the
origin, using an analytical continuation in this complex plain.
The relations between Heun functions of different values of
parameters and arguments, described in the mathematical literature
\cite{Heun}, are useful for this purpose, too.

Using these techniques we obtain the following local solutions of
the confluent Heun equation (\ref{H}):

1. Around the singular point $r=0$ there exist two independent
local solutions $H_{\,\,\varepsilon,s,l}^{(0)\pm}(r)$ with
asymptotics:
\ben H_{\,\,\varepsilon,s,l}^{(0)+}(r)\sim
{1\over{r^{2s}}},\,\,\,\,\,H_{\,\,\varepsilon,s,l}^{(0)-}(r)\sim
1.\la{as0}\een

The second one is the standard canonical solution:
\ben H_{\,\,\varepsilon,s,l}^{(0)-}(r)=HeunC\left(2i\varepsilon,
2s, 2i\varepsilon, 2\varepsilon^2,s^2-l(l+1),r
\right),\la{Sol1}\een
which has convergent Taylor series expansion around the origin
$r=0$. This solution is the only one, which is finite at the point
$r=0$. It is normalized according to the relations:
\begin{subequations}\label{norm1:abc}
\ben H_{\,\,\varepsilon,s,l}^{(0)-}(0)=HeunC\left(2i\varepsilon,
2s, 2i\varepsilon,
2\varepsilon^2,s^2-l(l+1),0 \right)=1, \la{norm1:a}\\
{d\over{dr}}H_{\,\,\varepsilon,s,l}^{(0)-}(0)={d\over{dr}}HeunC
\left(2i\varepsilon,2s, 2i\varepsilon, 2\varepsilon^2,s^2-l(l+1),0
\right)={{s(s+1)-l(l+1)}\over{2s+1}}.\la{norm1:b}\la{norm1:c}\een
\end{subequations}

In the general case, the solutions of this type diverge at the
horizon $r=1$. There exist a countable number of such solutions
which are finite simultaneously at the origin and at the horizon.
Considering them, one can derive a discrete spectrum of values
$\varepsilon_{n,s,l}$ for the boundary problem on the interval
$r\in [0,1]$. To the best of our knowledge, this boundary problem
has not attracted attention in the existing physical applications
and its spectrum is not known at present, although it must be
interesting for the study of the interior of Schwarzschild black
holes\footnote{Note that in this domain the variable $r$ plays the
role of a time variable and the dependence of the former
stationary solutions on the new space variable $t$ is described by
the simple exponential factor $e^{i\varepsilon t}$.}. Indeed,
these solutions are obviously square integrable functions on the
interval $r\in [0,1]$. If they will be proven to form a complete
set in the corresponding functional space, these solutions  will
present a natural basis of {\em normal modes} for a well defined
expansion of any small perturbation of the metric in the interior
of the black hole. Therefore these solutions seem to deserve a
separate consideration.

The first independent local solution in Eqs. (\ref{as0}) --
$H_{\,\,\varepsilon,s,l}^{(0)+}(r)$ cannot be described by
standard solution (\ref{Scan}) of the confluent Heun equation for
any set of parameters. Since $b=2s$ is positive integer, the
second solution includes the $\ln r$ terms \cite{Heun}. It is
analogous to the corresponding case of solutions of confluent
hypergeometric equation, but we were not able to find more
information about this solution in the available mathematical
literature. Taking into account the well-known example of hydrogen
atom, described by confluent hypergeometric functions, one can
think that this solution may not play an essential role in most of
the physical applications. Nevertheless, for completeness we shall
add formally this solution to our list, using for it the notation
$H_{\,\,\varepsilon,s,l}^{(0)+}(r)$.

2. We have two independent local solutions of the Heun confluent
equation (\ref{H}):
\begin{subequations}\label{Sol2:ab} \ben
 H_{\,\,\varepsilon,s,l}^{(1)+}(r)=HeunC\left(-2i\varepsilon,+ 2i\varepsilon,2s,
-2\varepsilon^2,2\varepsilon^2+s^2-l(l+1),1-r \right),\la{Sol2:a}\\
 H_{\,\,\varepsilon,s,l}^{(1)-}(r)=(r-1)^{-2i\varepsilon}HeunC\left(-2i\varepsilon,-
2i\varepsilon,2s, -2\varepsilon^2,2\varepsilon^2+s^2-l(l+1),1-r
\right).\la{Sol2:b}\een
\end{subequations}
The functions $H_{\,\,\varepsilon,s,l}^{(1)+}(r)$ and
$(r-1)^{2i\varepsilon}H_{\,\,\varepsilon,s,l}^{(1)-}(r)$ have
convergent Taylor series expansion around the second singular
point -- the horizon $r=1$. They are normalized in the following
way:
\begin{subequations}\label{NormSol2:ab} \ben
  \left. \!\!\begin{array}{cc}
H_{\,\,\varepsilon,s,l}^{(1)+}(r)\Big|_{{}_{r=1}}\,\\
 (r-1)^{2i\varepsilon}H_{\,\,\varepsilon,s,l}^{(1)-}(r)\Big|_{{}_{r=1}}
 \end{array}\!\!\right\}&=&1,\la{NormSol2:a}\\
 \left. \!\!\!\begin{array}{cc}
{d\over{dr}}H_{\,\,\varepsilon,s,l}^{(1)+}(r)\Big|_{{}_{r=1}}\,\\
{d\over{dr}}\!\left((r\!-\!1)^{2i\varepsilon}H_{\,\,\varepsilon,s,l}^{(1)-}(r)\right)\Big|_{\!{}_{r=1}}
\end{array}\!\!\!\right\}\!\!
 &=&\!\!
 {{2\varepsilon^2(1\mp 1)\pm 2i\varepsilon s\!+\!i\varepsilon(1\pm 1)\!+\!s(s\!+\!1)\!-\!l(l\!+\!1)}\over
 {\pm 2i\varepsilon+1}}.\hskip .8truecm\la{NormSol2:b}\een
\end{subequations}

The first solution (\ref{Sol2:a}) is the only one which is regular
at the horizon $r=1$. In general case, the two solutions
(\ref{Sol2:ab}) diverge at both the origin $r=0$ and the infinity
$r=\infty$. Using them one can formulate the corresponding
two-point boundary problems and find their discrete spectra. We
shall consider some of them in the next sections.

3. Using known three-term recurrence relations for the
coefficients $a_\nu^\pm$, one is able to construct two
asymptotics-series solutions around the infinite irregular
singular point $r=\infty$ \cite{Heun}:
\begin{subequations}\label{Sol3:ab}
\ben H_{\,\,\varepsilon,s,l}^{(\infty)+}(r)\sim
r^{-s-1}e^{-2i\varepsilon(r+\ln r)}\sum_{\nu\geq
0}{{a_\nu^{+}}\over{r^\nu}},\,\,\,\,a_0^+=1\la{Sol3:a}\\
H_{\,\,\varepsilon,s,l}^{(\infty)-}(r)\sim r^{-s-1}\sum_{\nu\geq
0}{{a_\nu^{-}}\over{r^\nu}},\,\,\,\,a_0^-=1; .\la{Sol3:b}\een
\end{subequations}
They are valid in the complex domain
\ben -{\pi\over 2}-l(l+1)+s^2\leq \arg(r)+\arg(\varepsilon)\leq
3\,{\pi\over 2}+l(l+1)-s^2, \la{domain}\een
see the articles by A.~Decarreau et al. in \cite{Heun}, where one
can find the restrictions of this kind for the general case of the
confluent Heun equation. In our particular problem Eq.
(\ref{domain}) leads to inessential restrictions on the variables
in the complex plain $\mathbb{C}$, since the length of the
admissible interval of arguments $\arg(r)$ and $\arg(\varepsilon)$
is $2\pi+2l(l+1)-2s^2\geq 2(\pi+s)\geq 2\pi$ for $l\geq s\geq 0$.

Unfortunately, the transition coefficients between the three sets
of local solutions (\ref{Sol1}), (\ref{Sol2:ab}) and
(\ref{Sol3:ab}) are not known at present. The only available
information is negative: these coefficients cannot be expressed
even in terms of hypergeometric functions \cite{Heun}. The same
transition coefficients relate the corresponding local solutions
of the Regge-Wheeler equation, see below.

\subsection{Local Solutions of Regge-Wheeler Equation}

In accord with the above local solutions of the Heun equation
(\ref{H}), we obtain the following three sets of independent local
solutions of the Regge-Wheeler equation (\ref{RW}):
\begin{subequations}\label{R0:abc}
\ben \Phi_{\,\,\varepsilon,s,l}^{(0)\pm}(t,r)=
e^{i \varepsilon t}R_{\,\,\varepsilon,s,l}^{(0)\pm}(r),
\hskip 8.9truecm \la{R0:a}\\
R_{\,\,\varepsilon,s,l}^{(0)\pm}(r)= r^{s+1} e^{ i\varepsilon r
+i\varepsilon\ln(1-r)} \begin{cases}
H_{\,\,\varepsilon,s,l}^{(0)+}(r),\,\\
HeunC\!\left(2i\varepsilon, 2s, 2i\varepsilon,
2\varepsilon^2,s^2-l(l+1),r \right) ;\end{cases}  \hskip
1.truecm\la{R0:b}\een
\end{subequations}
\begin{subequations}\label{R1:ab}
\ben \Phi_{\,\,\varepsilon,s,l}^{(1)\pm}(t,r)=
e^{i \varepsilon t}R_{\,\,\varepsilon,s,l}^{(1)\pm}(r),\hskip 8.9truecm \la{R1:a}\\
R_{\,\,\varepsilon,s,l}^{(1)\pm}(r)\!=\!r^{s+1} e^{i\varepsilon r
\pm i\varepsilon\ln(r-1)} HeunC\!\left(\!-2i\varepsilon,\!\pm
2i\varepsilon,2s,
\!-2\varepsilon^2,2\varepsilon^2\!+\!s^2\!-\!l(l\!+\!1),1\!-\!r\!
\right)\!;\la{R1:b}\een
\end{subequations}
%
\begin{subequations}\label{Rinf:ab}
\ben \Phi_{\,\,\varepsilon,s,l}^{(\infty)\pm}(t,r)=
e^{i \varepsilon t}R_{\,\,\varepsilon,s,l}^{(\infty)\pm}(r),\hskip 8.6truecm \la{Rinf:a}\\
   R_{\,\,\varepsilon,s,l}^{(\infty)\pm}(r)\sim{{e^{\mp i\varepsilon r -i\varepsilon\ln r}
}} \sum_{\nu\geq 0}{{a_\nu^{\pm}}\over{r^\nu}},\,\,\,\,a_0^\pm=1.
\hskip 6.3truecm\la{Rinf:b}\een
\end{subequations}
They are correspondingly defined in domains around the three
singular points of the problem: the origin $r=0$, horizon $r=1$
and infinity $r=\infty$, as pointed in the upper indices of our
notation. Any other radial function $R_{\varepsilon,s,l}(r)$,
being solution of the stationary Regge-Wheeler equation (\ref{R}),
can be represented locally in the form of a linear combination:
\ben R_{\varepsilon,s,l}(r)=
C_{X+}R_{\,\,\varepsilon,s,l}^{(X)+}(r)+
C_{X-}R_{\,\,\varepsilon,s,l}^{(X)-}(r),\la{linC}\een
where $C_{X\pm}$ are proper constants and $X=0,1,\infty$ indicates
the corresponding singular point.

For fixed values of the parameters $\varepsilon, s, l$ each of the
three sets (\ref{R0:abc}), (\ref{R1:ab}) and (\ref{Rinf:ab})
includes two independent local solutions of the stationary
Regge-Wheeler equation (\ref{R}). Hence, in the corresponding
common domains of validity of these solutions one obtains:
\ben R_{\,\,\varepsilon,s,l}^{(X)\pm}(r)=
\Gamma_{\,Y+}^{X\pm}(\varepsilon,s,l)
R_{\,\,\varepsilon,s,l}^{(Y)+}(r)+
\Gamma_{\,Y-}^{X\pm}(\varepsilon,s,l)
R_{\,\,\varepsilon,s,l}^{(Y)-}(r),\la{RGamma}\een
expanding one set of local solutions onto another. The main
obstacle for analytical treatment of different interesting
problems related to the Regge-Wheeler equation is that the
transition coefficients $\Gamma_{\,Y\pm}^{X\pm}(\varepsilon,s,l)$
for $X,Y=0,1,\infty$; $X\neq Y$ are unknown explicitly.

\subsection{Some Basic Properties of the Local Solutions of Regge-Wheeler Equation}

\subsubsection{In-Out Properties of the Local Solutions}

Taking into account the relation (\ref{norm:a}), one easily
obtains directions of spreading of the corresponding waves
described by local solutions (\ref{R0:a}), (\ref{R1:a}) and
(\ref{Rinf:a}) in the vicinity of the corresponding singular
points $X$ when the time $t$ increases\footnote{The only still
unstudied solution $H_{\,\,\varepsilon,s,l}^{(0)+}(r)$ has to be
chosen in a proper way to insure the fulfilment of our "in-out"
conventions. This requirement fixes the solution
$H_{\,\,\varepsilon,s,l}^{(0)+}(r)$ up to inessential
normalization.}. These "in" and "out" properties of local
solutions are illustrated by the graphical schemes, shown in
Table~\ref{table1}:
\vskip 1.truecm
\begin{table}[here]~\vspace{-1.truecm}
\begin{center}
\begin{tabular}{cllllrrrr}
\hline
$\Phi_{\,\,\varepsilon,s,l}^{(0)+}(t,r):$&$\,\,\,0\leftarrow$&$\Big|
$&$\Phi_{\,\,\varepsilon,s,l}^{(0)-}(t,r):$&$\,\,\,0\rightarrow$&\\
\hline
$\Phi_{\,\,\varepsilon,s,l}^{(1)+}(t,r):$&$\,\,\,1\leftarrow$&$\Big|
$&$\Phi_{\,\,\varepsilon,s,l}^{(1)-}(t,r):$&$\,\,\,1\rightarrow$&\\
\hline
$\Phi_{\,\,\varepsilon,s,l}^{(\infty)+}(t,r):$&$\,\,\,\,\rightarrow\infty$&$\Big|
$&$\Phi_{\,\,\varepsilon,s,l}^{(\infty)-}(t,r):$&$\,\,\,\,\leftarrow\infty$&\\
\hline
\end{tabular}
\end{center}
\caption{In-Out properties of local solutions}
\la{table1}\end{table}
\vskip .4truecm

Thus, we see that according to our conventions, the local
solutions $\Phi_{\,\,\varepsilon,s,l}^{(X)+}$ describe waves going
in the singular point $X$ and the local solutions
$\Phi_{\,\,\varepsilon,s,l}^{(X)-}$ describe waves going out of
singular point $X$.

\subsubsection{Time and Space Limits of Local Solutions for
Complex Values of $\varepsilon$}

In general, we have to consider complex valued
$\varepsilon=\varepsilon_R+i\varepsilon_I\in
\mathbb{C}_\varepsilon$. Since in the Regge-Wheeler equation only
the squared quantity $\varepsilon$ emerges, without loss of
generality one can accept conventionally that its real part is
nonnegative: $\varepsilon_R \geq 0$ \footnote{In the case
$\varepsilon_R = 0$, we may have "pseudo" waves. The corresponding
solutions do not oscillate.}.

The local solutions (\ref{R0:a}), (\ref{R1:a}) and (\ref{Rinf:a})
have the following obvious time-limits:
\ben \lim_{t\to +\infty}\la{t_lim}
|\Phi_{\,\,\varepsilon,s,l}^{(X)\pm}(t,r)|=\begin{cases} 0\,\,\,
\,\,\forall r, &\hbox{if}\,\,\,
\varepsilon_I>0\,\,\,\hbox{and}\,\,\,|R_{\,\,\varepsilon,s,l}^{(X)\pm}(r)|\neq
\infty ;\\|R_{\,\,\varepsilon,s,l}^{(X)\pm}(r)|, &\hbox{if}\,\,\,
\varepsilon_I=0;
\\\infty\,\,\,\forall r, & \hbox{if}\,\,\,\varepsilon_I <0
\,\,\,\hbox{and}\,\,\,|R_{\,\,\varepsilon,s,l}^{(X)\pm}(r)|\neq 0.
\end{cases}\een

Hence, from a physical point of view we have to work only with
$\varepsilon_I \geq 0$ \cite{RW}, because the solutions with
$\varepsilon_I < 0$ are certainly unstable and, therefore,
nonphysical. Then, for finite values of $t$ in general we obtain
the following space limits of the local solutions:
\begin{subequations}\label{r_lim:ab}
\ben \lim_{r\to X}
|\Phi_{\,\,\varepsilon,s,l}^{(X)+}(t,r)|&=&\infty,\label{r_lim:a}\\
\lim_{r\to X} |\Phi_{\,\,\varepsilon,s,l}^{(X)-}(t,r)|&=&0.
\label{r_lim:b}\een
\end{subequations}
These need some justification for different singular points:

1. The limits (\ref{r_lim:ab}) of local solutions around the
origin $r\!=\!0$ are valid for $s\!>\!0$ and on any contour to
this point in the complex plane $\mathbb{C}_r$. For $s\!=\!0$ the
limit (\ref{r_lim:b}) equals $1$.

2. Around the horizon $r=1$ the limits (\ref{r_lim:ab}) of local
solutions are valid for all real values of $s\geq 0$ and on any
contour to this point in the complex plane $\mathbb{C}_r$.

3. The behavior of limits (\ref{r_lim:ab}) of local solutions in
the complex plane $\mathbb{C}_r$ around the infinite irregular
singular point $r=\infty$ is more complicated and depends on the
direction of the contour which approaches this point. This
reflects the well-known Stocks phenomenon (see references in
\cite{QNM, Heun}), and is related with the multiplier
$$e^{\mp i\varepsilon r
-i\varepsilon\ln r}=e^{i\left(\alpha\mp
|\varepsilon||r|\cos(\arg(r)+\arg(\varepsilon))\right)}
e^{\pm|\varepsilon||r|\sin(\arg(r)+\arg(\varepsilon))+
|\varepsilon|\ln|r|\cos(\arg(\varepsilon))}$$
in (\ref{Rinf:b}). Thus we see that
\ben |R_{\,\,\varepsilon,s,l}^{(\infty)\pm}(r)|\sim
{{e^{\pm|\varepsilon||r|\sin(\arg(r)+\arg(\varepsilon))+
|\varepsilon|\ln|r|\cos(\arg(\varepsilon))}}}, \la{R_asimptotic}\een
and in the case $|\varepsilon|>0$ we obtain:
\begin{subequations}\label{limRinf:ab}
\ben \lim_{|r|\to \infty}
|R_{\,\,\varepsilon,s,l}^{(\infty)+}(r)|=\begin{cases}
\infty,\,\,\, &\hbox{if}\,\,\,
\arg(r)+\arg(\varepsilon)\in(0,\pi)/
\hskip -.4truecm\mod 2\pi;\label{limRinf:a}\\
0, \,\,\, &\hbox{if}\,\,\,
\arg(r)+\arg(\varepsilon)\in(-\pi,0)/\hskip -.4truecm\mod 2\pi.
\end{cases}\\
\lim_{|r|\to \infty}
|R_{\,\,\varepsilon,s,l}^{(\infty)-}(r)|=\begin{cases} 0,\,\,\,
&\hbox{if}\,\,\, \arg(r)+\arg(\varepsilon)\in(0,\pi)/
\hskip -.4truecm\mod 2\pi;\label{limRinf:b}\\
\infty, \,\,\, &\hbox{if}\,\,\,
\arg(r)+\arg(\varepsilon)\in(-\pi,0)/\hskip -.4truecm\mod 2\pi.
\end{cases}
\een \end{subequations}

In the case $|\varepsilon|=0$ we obtain:
\ben \lim_{|r|\to \infty}
|R_{\,\,\varepsilon,s,l}^{(\infty)\pm}(r)|=1.\la{limRinf0}\een

Relations (\ref{limRinf:ab}) and (\ref{limRinf0}) justify the
exact behavior of local solutions (\ref{Rinf:ab}), when one
approaches the infinite singular point from different directions
in the complex plain $\mathbb{C}_r$. They give precise meaning of
limits (\ref{r_lim:ab}) for $X=\infty$. These relations will play
an important role in our novel approach to different spectra of
the corresponding two-points boundary problems related to the
Regge-Wheeler equation.

\section{Two-Point Boundary Problems}

\subsection{General Description and Symbols of Two-Point Boundary
Problems for Regge-Wheeler Equation}

The two-singular-point boundary problems can be defined on the
{\em real} intervals which contain only two singular end-points
and have no other singular points inside. It is possible to
consider regular-singular two-point boundary problems too. In this
case, one end of the interval is a singular point and the another
is a regular point supplied with some standard boundary condition
(for example -- with Dirichlet's boundary condition, see below).
Different boundary conditions correspond to different physical
problems described by the same differential equation.

One can find the general definition of a stationary
two-singular-point boundary problem in the book by Slavyanov and
Lay in \cite{Heun}. For our problem we need some natural
generalization of this definition which includes not only the case
of solutions, which are finite at both singular points. Following
the physical traditions (see articles \cite{QNM}), we use as a
basis of such a definition the in-out properties of the
time-dependent solutions of the Regge-Wheeler equation, considered
in the previous Section. Since we have establish the one-to-one
correspondence between these properties and the corresponding
space limits (\ref{r_lim:ab}) on the {\em real} axes
$\mathbb{R}_r$, we can reformulate the definition using only the
solutions of the stationary Regge-Wheeler equation (\ref{R}) in
the spirit of the book by Slavyanov and Lay.

All two-points boundary problems can be formulated using proper
constraints on the transition coefficients
$\Gamma_{Y\pm}^{X\pm}(\varepsilon,s,l)$ in relations
(\ref{RGamma}). This way one obtains equations for finding of
corresponding discrete spectra of (possibly complex) values of
frequencies $\varepsilon$.

\subsection{Two-Singular-Point Problems on the interval $(1,\infty)$}

This interval corresponds to the outer domain of Schwarzschild
black holes, where the eventual observer lives. Therefore, it is
most important from a physical point of view and, correspondingly,
most well studied at present. Nevertheless, even for this interval
much further work remains to be done.

\subsubsection{Quasi-Normal Modes}
We start the illustration of the general statements, made in
previous Section, with the most well studied at present example of
quasi-normal modes (QNM) of Schwarzschild black holes \cite{QNM},
as a specific two-singular-point boundary problem on the interval
$(1,\infty)$. Using the graphical scheme, shown in Table
\ref{table1}, we can symbolically write down relation
(\ref{RGamma}) in terms of time dependent solutions of the
Regge-Wheeler equation as follows:
\ben \{1\leftarrow\}= \Gamma_{\infty +}^{\,1\,\,
+}(\varepsilon,s,l) \{\rightarrow\infty\}+ \Gamma_{\infty
-}^{\,1\,\, +}(\varepsilon,s,l) \{\leftarrow\infty\}.
\la{1_left_inf}\een

By definition, for QNM we have no outgoing waves from the infinite
point \cite{QNM}, i.e.
\ben \Gamma_{\infty -}^{\,1\,\,+}(\varepsilon,s,l)=0\,\,\,
\Rightarrow \varepsilon=\varepsilon(n,s,l),\,\,\,n=0,1,2,\dots.
\la{QNMspectrum}\een
This equation generates the spectrum of QNM. For the above values
of $\varepsilon$  Eq. (\ref{RGamma}) reads: \
\ben R_{\,\,\varepsilon(n,s,l),s,l}^{(1)+}(r)= \Gamma_{\infty
+}^{\,1\,\,+}(\varepsilon(n,s,l),s,l)
R_{\,\,\varepsilon(n,s,l),s,l}^{(\infty)+}(r).\la{RQNM}\een

We can simplify significantly the notation if we introduce the
symbol "$1\!\smile\!\infty$" for this two-point boundary problem.
It reflects symbolically the properties (\ref{r_lim:ab}) of its
solutions which enter into relation (\ref{RQNM}). Accordingly, we
denote by $\Phi_{1\smile
\infty}(t,r|{n,s,l})=e^{i\varepsilon_{1\smile \infty}({n,s,l})
t}R_{1\smile \infty}(r|{n,s,l})$ and $R_{1\smile
\infty}(r|{n,s,l})=R_{\,\,\varepsilon(n,s,l),s,l}^{(1)+}(r)$ the
solutions of this boundary problem, and by $\varepsilon_{1\smile
\infty}({n,s,l})$, the corresponding eigenvalues.

\subsubsection{Left Mixed Modes}
Under the condition
\ben \Gamma_{\infty +}^{\,1\,\,+}(\varepsilon,s,l)=0\,\,\,
\Rightarrow \varepsilon=\varepsilon_{1\backsim\,
\infty}(n,s,l),\,\,\,n=0,1,2,\dots; \la{LMMspectrum}\een
we obtain the spectrum of the waves, going left from infinity to
horizon $r=1$  without reflection -- left mixed modes
(LMM)\footnote{We use the term "mixed modes" for indication of the
following obvious property of these solutions: Their boundary
conditions at the two end points of the corresponding interval are
of different type, in the sense of Eq. (\ref{r_lim:ab}).}. Due to
Eq. (\ref{r_lim:ab}) we use for them the symbol
$1\!\backsim\!\infty$.

Considering the equation
\ben \{1\rightarrow\}= \Gamma_{\infty +}^{\,1\,\,
-}(\varepsilon,s,l) \{\rightarrow\infty\}+ \Gamma_{\infty
-}^{\,1\,\, -}(\varepsilon,s,l) \{\leftarrow\infty\}
\la{1_right_inf}\een
we can introduce two more types of modes of Schwarzschild black
holes:

\subsubsection{Normal Modes}
The equation
\ben \Gamma_{\infty +}^{\,1\,\,-}(\varepsilon,s,l)=0\,\,\,
\Rightarrow
\varepsilon=\varepsilon_{1\frown\,\infty}(n,s,l),\,\,\,n=0,1,2,\dots;
\la{NMspectrum}\een
defines the spectrum of normal modes (NM) on the interval
$(1,\infty)$, which are important for considerations of the
stability of Schwarzschild black holes. For obvious reasons,
related once more to Eq. (\ref{r_lim:ab}),  we use for them the
symbol $1\!\frown\!\infty$.

\subsubsection{Right Mixed Modes}
Finally, the equation
\ben \Gamma_{\infty -}^{\,1\,\,-}(\varepsilon,s,l)=0\,\,\,
\Rightarrow
\varepsilon=\varepsilon_{1\thicksim\,\infty}(n,s,l),\,\,\,n=0,1,2,\dots;
\la{RMMspectrum}\een
defines the spectrum of the waves, going right from the horizon
$r=1$ to infinity without reflection -- right mixed modes (RMM).
Due to Eq. (\ref{r_lim:ab}) we use for them the symbol
$1\!\thicksim\!\infty$.

\subsection{Two-Singular-Point Problems on the intervals $(X,Y)$}

The generalization of the previous consideration for any real
interval $(X,Y)$ with singular ends $X<Y$ and without other
singular points in it is obvious. One can introduce the
corresponding:

a) Quasi-normal modes: $\Phi_{X\smile Y}(t,r|{n,s,l})
=e^{i\varepsilon_{X\smile Y}({n,s,l}) t}R_{X\smile Y}(r|{n,s,l})$
-- the solutions of the boundary problem with eigenvalues
$\varepsilon_{X\smile Y}({n,s,l})$ defined by the equation:
\ben \Gamma_{Y -}^{X +}(\varepsilon,s,l)=0\,\,\, \Rightarrow
\varepsilon=\varepsilon_{X\smile
Y}({n,s,l}),\,\,\,n=0,1,2,\dots\,. \la{XYQNMspectrum}\een

b) Left mixed modes: $\Phi_{X\backsim Y}(t,r|{n,s,l})
=e^{i\varepsilon_{X\backsim Y}({n,s,l}) t}R_{X\backsim
Y}(r|{n,s,l})$ -- the solutions with eigenvalues
$\varepsilon_{X\backsim Y}({n,s,l})$ defined by the equation:
\ben \Gamma_{Y +}^{X +}(\varepsilon,s,l)=0\,\,\, \Rightarrow
\varepsilon=\varepsilon_{X\backsim\,
Y}(n,s,l),\,\,\,n=0,1,2,\dots\,. \la{XYLMMspectrum}\een

c) Normal Modes: $\Phi_{X\frown Y}(t,r|{n,s,l})
=e^{i\varepsilon_{X\frown Y}({n,s,l}) t}R_{X\frown Y}(r|{n,s,l})$
-- the solutions with eigenvalues $\varepsilon_{X\frown
Y}({n,s,l})$ defined by the equation:
\ben \Gamma_{Y +}^{X -}(\varepsilon,s,l)=0\,\,\, \Rightarrow
\varepsilon=\varepsilon_{X\frown Y}(n,s,l),\,\,\,n=0,1,2,\dots\,.
\la{XYNMspectrum}\een

d) Right mixed modes: $\Phi_{X\thicksim Y}(t,r|{n,s,l})
=e^{i\varepsilon_{X\thicksim Y}({n,s,l}) t}R_{X\thicksim
Y}(r|{n,s,l})$ -- the solutions with eigenvalues
$\varepsilon_{X\thicksim Y}({n,s,l})$, defined by equation:
\ben \Gamma_{Y -}^{X -}(\varepsilon,s,l)=0\,\,\, \Rightarrow
\varepsilon=\varepsilon_{X\thicksim
Y}(n,s,l),\,\,\,n=0,1,2,\dots\,. \la{XYRMMspectrum}\een

To the best of our knowledge, all spectra of Schwarzschild black
hole two-point boundary problems on the intervals $(X,Y)\neq
(1,\infty)$ at present are unknown both analytically and
numerically. Much better is the situation in the case of the
interval $(1,\infty)$: the spectrum of QNM is known in detail,
both numerically and analytically \cite{QNM}, although no exact
explicit expressions have been obtained up to now. In addition, in
this case we have some qualitative studies of the spectrum of
normal modes \cite{NM}.

\subsection{Regular-Singular-Two-Point Boundary Problems}

\subsubsection{Formulation of the Problem and Notations}
These two-point boundary problems for the Regge-Wheeler equation
(\ref{RW}) are related to models of heavy compact objects which
are essentially different from a black hole model. The problem of
finding the corresponding spectra in the {\em simple} mathematical
formulation, considered below, seems to be new and even not posed
in the existing physical literature.

In the present section we consider a one-parameter family of real
intervals $(r_{\!*}, \infty)\subset (1,\infty)$. The left end of
the interval is an arbitrary {\em regular} point $r_{\!*}\in
(1,\infty)$ of Eq. (\ref{R}). We pose Dirichlet's boundary
condition at this end:
\ben \Phi_{\varepsilon,s,l}(t,r_{\!*})=0.\la{Dirichlet}\een

Physically, this means that the waves are experiencing a {\em
total} reflection at the spherical surface with the area radius
$r=r_{\!*}>1$, instead of going freely through it without any
reflection, as in the case of black holes. We have to mention that
Dirichlet's boundary condition at the cosmological horizon has
been used recently for QNM of AdS black holes in the article
\cite{Horowitz&Hubeny}.

\begin{figure}[htbp] \vspace{6.truecm}
\includegraphics{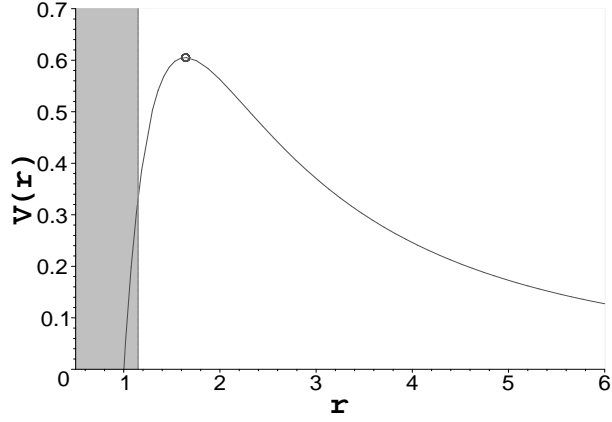} \caption{\hskip 0.2truecm The potential $V_{s,l}(r)$
for $s=2,\,\,l=2$. The small circle marks the place of its maximum
(here and at all figures hereafter). The shadowed domain marks the
area, which is forbidden for spreading of waves. Its right edge is
at the point $r_*$. The point $r=1$ presents the event horizon.
    \hskip 1truecm}
    \label{Fig1}
\end{figure}

One can consider this surface as a boundary of some {\em massive}
spherically symmetric body.  Owing to the Birkhoff theorem, in
general relativity the gravitational field of any massive body
with Keplerian mass $M$ outside the radius $r_{\!*}$ coincides
with the corresponding field of a black hole. Hence, the spreading
of the waves in the outer domain of the massive body is governed
by the Regge-Wheeler equation (\ref{RW}), too, but the boundary
conditions must be different. The corresponding effective
potential $V_{s,l}(r)$ in the Regge-Wheeler equation (\ref{RW}) is
illustrated in Fig. \ref{Fig1}.

There exists a large amount of articles on QNM of relativistic
stars \cite{QNM, QNMinStars}, both stationary and rotating. At
present, this field is well studied in detail. It is known that,
in general, the interior of the stars may play a complicated role
in the polar oscillations of fields of different spin outside the
star. Some details are still an open problem, because the
relativistic equation of state of matter is not precisely known
for the real case of very high pressures in the compact
relativistic stars, for example, in neutron stars.

For axial perturbations of metric the situation is different:
axial perturbations do not induce a motion of the star matter and
experience only a potential scattering, as in the case of
Schwarzschild black holes \cite{QNM}. This is a physical reason
for our simple model of axial perturbations of gravitational field
of compact objects.

In this model, we avoid all complicated problems related to the
body's interior, replacing them with the simple boundary condition
(\ref{Dirichlet}). It can be thought as idealization, which
describes a proper limiting case of problems related to real
bodies. It is obvious that this idealization is opposite to the
Schwarzschild black hole model. If there exist a conserved current
in the problem at hand, one can expect that the corresponding
eigenvalues for real static spherically symmetric objects will be
placed between the corresponding eigenvalues of black holes and
the eigenvalues of the simple model of massive compact objects, we
study here for the first time. Unfortunately, such simple guess
turns to be incorrect, because the current in the RW problem with
complex eigenvalues $\varepsilon$ is not conserved and has quite
nontrivial properties, see the Appendix.

In general, one can try to consider two different boundary
problems:

a) Quasi-normal modes of massive compact objects with the spectrum
$\varepsilon_{ r_{\!*}\,\bullet\hskip -0.1truecm\thicksim\, \infty
}(n,s,l)$: $\Phi_{r_{\!*}\,\bullet\hskip -0.1truecm\thicksim\,
\infty}(t,r|n,s,l)= e^{i\varepsilon_{r_{\!*}\,\bullet\hskip
-0.1truecm\thicksim\, \infty}(n,s,l) t}R_{r_{\!*}\,\bullet\hskip
-0.1truecm\thicksim\, \infty}(r|n,s,l)$, $n=0,1,2,\dots$ .

b) Normal modes of massive compact objects with the spectrum
$\varepsilon_{ r_{\!*}\,\bullet\hskip -0.1truecm\backsim\, \infty
}(n,s,l)$: $\Phi_{r_{\!*}\,\bullet\hskip -0.1truecm\backsim\,
\infty}(t,r|n,s,l)= e^{i\varepsilon_{r_{\!*}\,\bullet\hskip
-0.1truecm\backsim\, \infty}(n,s,l) t}R_{r_{\!*}\,\bullet\hskip
-0.1truecm\backsim\, \infty}(r|n,s,l)$, $n=0,1,2,\dots$ .

Here we use the symbols $r_{\!*} \,\hbox{$\bullet\hskip
-.23truecm\thicksim$} \infty$ and $ r_{\!*} \,\hbox{$\bullet\hskip
-.23truecm\backsim$} \infty$ for QNM and NM in the exterior domain
of compact massive objects. In these symbols the sign "$r_{\!*}
\,\bullet$" denotes the spherically symmetric compact object with
the boundary of the area radius $r_{\!*}$. The tail denotes the
corresponding perturbation mode in accord with the properties
described by Eq. (\ref{r_lim:ab})\footnote{From a pure
mathematical point of view it is possible to consider analogous
modes on intervals $(r_{\!*},X)$ with $X=1, r_{\!*}\in (0,1)$ and
$X=1, r_{\!*}\in (1,\infty)$, or $X=0, r_{\!*}\in(0,1)$ and even
$X=0, r_{\!*}\in(-\infty,0)$, using symbols like $r_{\!*}
\,\hbox{$\bullet\hskip -.23truecm\thicksim$} X$ and $ r_{\!*}
\,\hbox{$\bullet\hskip -.23truecm\backsim$} X$ for the
corresponding QNM and NM. Since the physical meaning of such modes
is not clear, we will ignore them in the present article.}.

Using the properties of the complex current in RW problem one can
prove that NM do not exist under Dirichlet's boundary condition
(\ref{Dirichlet}), see the Appendix. This is well understandable
from physical point of view, since in the problem at hand we have
no absorbtion of the waves, but a total reflection, creation and
emission to space infinity.

\subsubsection{The Equation for the Discrete Spectrum of QNM}

Let us consider solutions of the Regge-Wheeler equation
$\Phi_{\varepsilon,s,l}^{\,\,\,r_{\!*}}(t,r)$ on the interval
$(r_*,\infty)$, which obey the boundary condition
(\ref{Dirichlet}). Making use of solutions (\ref{R1:b}) of the
stationary equation (\ref{R}), we can represent the solutions
$\Phi_{\varepsilon,s,l}^{\,\,\,r_{\!*}}(t,r)$ in the following
determinant-form:
\ben \Phi_{\varepsilon,s,l}^{\,\,\,r_{\!*}}(t,r)\!=\!
e^{i\varepsilon t} \begin{vmatrix}
R_{\,\,\varepsilon,s,l}^{(1)+}(r) &
R_{\,\,\varepsilon,s,l}^{(1)-}(r)\\
R_{\,\,\varepsilon,s,l}^{(1)+}(r_{\!*}) &
R_{\,\,\varepsilon,s,l}^{(1)-}(r_{\!*})
\end{vmatrix}\!=\!
R_{\,\,\varepsilon,s,l}^{(1)-}(r_{\!*})\Phi_{\varepsilon,s,l}^{(1)+}(t,r)\!-\!
R_{\,\,\varepsilon,s,l}^{(1)+}(r_{\!*})\Phi_{\varepsilon,s,l}^{(1)-}(t,r).\,\,
\la{SolutionDirichlet}\een

Their expansion on the basis of solutions (\ref{Rinf:ab}),
represented in the symbolic form reads:
\ben \Phi_{\varepsilon,s,l}^{\,\,\,r_{\!*}}(t,r)= \Gamma_{\infty
+}^{\,\,\,r_{\!*}}(\varepsilon,s,l) \{\rightarrow\infty\}+
\Gamma_{\infty -}^{\,\,\,r_{\!*}}(\varepsilon,s,l)
\{\leftarrow\infty\}, \la{r_*_inf}\een
where $\Gamma_{\infty \pm}^{\,\,\,r_{\!*}}(\varepsilon,s,l)$ are
proper coefficients. It is obvious that the spectrum of QNM is
defined by the following equation:
\ben \Gamma_{\infty -}^{\,\,\,r_{\!*}}(\varepsilon,s,l)=0\,\,\,
\Rightarrow \varepsilon_{ r_{\!*}\,\bullet\hskip
-0.1truecm\thicksim\, \infty }(n,s,l),\,\,\,n=0,1,2,\dots\,.
\la{r*QNMspectrum}\een

Now one can use relations (\ref{linC}) and (\ref{RGamma}) to
obtain:
\ben  \Gamma_{\infty \pm}^{\,\,\,r_{\!*}}(\varepsilon,s,l)=
\Gamma_{\infty \pm}^{\,1\,\,+}(\varepsilon,s,l)
R_{\,\,\varepsilon,s,l}^{(1)-}(r_{\!*})- \Gamma_{\infty
\pm}^{\,1\,\,-}(\varepsilon,s,l)
R_{\,\,\varepsilon,s,l}^{(1)+}(r_{\!*})\la{GG}\een
and to represent Eq. (\ref{r*QNMspectrum}) in the form:
\ben {{\Gamma_{\infty
-}^{\,1\,\,+}(\varepsilon,s,l)}\over{\Gamma_{\infty
-}^{\,1\,\,-}(\varepsilon,s,l)}}=
{{R_{\,\,\varepsilon,s,l}^{(1)+}(r_{\!*})}\over
{R_{\,\,\varepsilon,s,l}^{(1)-}(r_{\!*})}}=\hskip 3.truecm\nonumber \\
e^{2i\varepsilon\ln(r_{\!*}-1)} {{
HeunC\!\left(\!-2i\varepsilon,\!+ 2i\varepsilon,2s,
\!-2\varepsilon^2,2\varepsilon^2\!+\!s^2\!-\!l(l\!+\!1),1\!-\!r_{\!*}\!
\right)}\over{ HeunC\!\left(\!-2i\varepsilon,\!- 2i\varepsilon,2s,
\!-2\varepsilon^2,2\varepsilon^2\!+\!s^2\!-\!l(l\!+\!1),1\!-\!r_{\!*}\!
\right)}}.\la{r*spectra}\een
Here we have used the relations (\ref{R1:b}), too. Relation
(\ref{r*spectra}) present the final form of the equation for the
spectrum of QNM in our simple model of massive compact bodies.

\subsubsection{Some General Properties of Trajectories of
QNM Eigenvalues $\varepsilon_{ r_{\!*}\,\bullet\hskip
-0.1truecm\thicksim\, \infty }(n,s,l)$ in the Complex Plane
$\mathbb{C}_{\varepsilon_{ r_{\!*}\bullet\hskip
-0.1truecm\thicksim\, \infty } }$}

For the massive compact bodies with different values of the area
radius of their surface the eigenvalue $\varepsilon_{
r_{\!*}\,\bullet\hskip -0.1truecm\thicksim\, \infty }(n,s,l)$, as
a function of $ r_{\!*}\in (1,\infty)$, runs on some trajectory in
the complex plane $\mathbb{C}_{\varepsilon_{ r_{\!*}\bullet\hskip
-0.1truecm\thicksim\,\infty}}$. Here we prove three simple
properties of these trajectories. These properties are confirmed
in the next section by numerical analysis of trajectories and
illustrated in Figs. (\ref{Fig2}) and (\ref{Fig3}).

\vskip .3truecm

{\bf Proposition 1}: {\em If the limit  $r_{\!*}\to 1+0$ of the
trajectory of $\varepsilon_{ r_{\!*}\,\bullet\hskip
-0.1truecm\thicksim\, \infty }(n,s,l)$ in the complex plane
$\mathbb{C}_{ {\varepsilon_{ r_{\!*}\bullet\hskip
-0.1truecm\thicksim\, \infty } }}$ exist, then we have:}
\ben \lim_{r_{\!*}\to 1+0} \varepsilon_{r_{\!*}\,\bullet\hskip
-0.1truecm\thicksim\, \infty }(n,s,l)=0.\la{r*to0}\een
{\bf Proof:} The eigenvalue $\varepsilon_{r_{\!*}\,\bullet\hskip
-0.1truecm\thicksim\, \infty }(n,s,l)$ is obtained from the
corresponding Eq. (\ref{r*spectra}). Hence, its dependence on
$r_{\!*}$ is defined by the rhs of this equation.

The lhs of Eq. (\ref{r*spectra}) must be nonzero and finite,
because the solutions (\ref{R1:ab}) and (\ref{Rinf:ab}) are two
complete sets of independent local solutions of the Regge-Wheeler
equation.

Suppose that the limit (\ref{r*to0}) exists. According to the
relations (\ref{NormSol2:a}) and (\ref{R1:b}), the limit
$r_{\!*}\to 1+0$ of the rhs of Eq. (\ref{r*spectra}):
\ben \lim_{r_{\!*}\to
1+0}e^{2i\varepsilon\ln(r_{\!*}-1)}=\lim_{r_{\!*}\to
1+0}\left(e^{2i\Re(\varepsilon_{r_{\!*}\,\bullet\hskip
-0.1truecm\thicksim\, \infty
}(n,s,l))\ln(r_{\!*}-1)}e^{-2\Im(\varepsilon_{r_{\!*}\,\bullet\hskip
-0.1truecm\thicksim\, \infty
}(n,s,l))\ln(r_{\!*}-1)}\right)\la{lim}\een
does not exist if  $\lim_{r_{\!*}\to
1+0}\left(\Re(\varepsilon_{r_{\!*}\,\bullet\hskip
-0.1truecm\thicksim\, \infty }(n,s,l))\right)\neq 0$. If
$\lim_{r_{\!*}\to
1+0}\left(\Re(\varepsilon_{r_{\!*}\,\bullet\hskip
-0.1truecm\thicksim\, \infty }(n,s,l))\right)= 0$ and
$\lim_{r_{\!*}\to
1+0}\left(\Im(\varepsilon_{r_{\!*}\,\bullet\hskip
-0.1truecm\thicksim\, \infty }(n,s,l))\right)> 0$, the limit
(\ref{lim}) is infinite. In the case $\lim_{r_{\!*}\to
1+0}\left(\Im(\varepsilon_{r_{\!*}\,\bullet\hskip
-0.1truecm\thicksim\, \infty }(n,s,l))\right)< 0$ the limit
(\ref{lim}) is zero. Hence, if a nonzero finite limit (\ref{lim})
exists, Eq. (\ref{r*to0}) must be fulfilled.$\diamond$\vskip
.3truecm

{\bf Proposition 2}: {\em In the limit  $r_{\!*}\to +\infty$ on
the trajectory of $\varepsilon_{ r_{\!*}\,\bullet\hskip
-0.1truecm\thicksim\, \infty }(n,s,l)$ in the complex plane
$\mathbb{C}_{{\varepsilon_{ r_{\!*}\bullet\hskip
-0.1truecm\thicksim\, \infty } }}$ we have:}
\ben \lim_{r_{\!*}\to +\infty} \varepsilon_{r_{\!*}\,\bullet\hskip
-0.1truecm\thicksim\, \infty }(n,s,l)=0.\la{r*toInf}\een
{\bf Proof:} Using the expansion on
$R_{\,\,\varepsilon,s,l}^{(\infty)\pm}(r_{\!*})$ of type
(\ref{RGamma}) for $R_{\,\,\varepsilon,s,l}^{(1)\pm}(r_{\!*})$ in
Eq. (\ref{r*spectra}) and the properties (\ref{limRinf:ab}), one
easily obtains the limit $r_{\!*}\to +\infty$ of both sides of
relation (\ref{r*spectra}) in the form:
\ben \lim_{r_{\!*}\to +\infty}\left( {{\Gamma_{\infty
-}^{\,1\,\,+}(\varepsilon_{ r_{\!*}\,\bullet\hskip
-0.1truecm\thicksim\, \infty }(n,s,l),s,l)}\over{\Gamma_{\infty
-}^{\,1\,\,-}(\varepsilon_{ r_{\!*}\,\bullet\hskip
-0.1truecm\thicksim\, \infty
}(n,s,l),s,l)}}\right)=\lim_{r_{\!*}\to +\infty}
\left({{\Gamma_{\infty +}^{\,1\,\,+}(\varepsilon_{
r_{\!*}\,\bullet\hskip -0.1truecm\thicksim\, \infty
}(n,s,l),s,l)}\over{\Gamma_{\infty +}^{\,1\,\,-}(\varepsilon_{
r_{\!*}\,\bullet\hskip -0.1truecm\thicksim\, \infty
}(n,s,l),s,l)}}\right). \la{Inf_r*spectra}\een

This means that  in the limit $r_{\!*}\to +\infty$ the solutions
$R_{\,\,\varepsilon,s,l}^{(1)\pm}(r)$ become linearly dependent.
Formulae (\ref{R1:ab}) show that this is possible only when
equation (\ref{r*toInf}) takes place. $\diamond$\vskip .3truecm

{\bf Proposition 3}: {\em The trajectory of eigenvalue
$\varepsilon_{ r_{\!*}\,\bullet\hskip -0.1truecm\thicksim\, \infty
}(n,s,l)$ in the complex plane $\mathbb{C}_{{\varepsilon_{
r_{\!*}\bullet\hskip -0.1truecm\thicksim\, \infty } }}$ has no
cusps, i.e. the real part $\Re\left(\varepsilon_{
r_{\!*}\,\bullet\hskip -0.1truecm\thicksim\, \infty
}(n,s,l)\right)$ and the imaginary part $\Im\left(\varepsilon_{
r_{\!*}\,\bullet\hskip -0.1truecm\thicksim\, \infty
}(n,s,l)\right)$ can not have simultaneously extremum for finite
values of $ r_{\!*}\in (1,\infty)$}.

{\bf Proof:} Let us introduce the short notation
$g(\varepsilon)={{\Gamma_{\infty
-}^{\,1\,\,+}(\varepsilon,s,l)}\big/ {\Gamma_{\infty -
}^{\,1\,\,-}(\varepsilon,s,l)}}$ and
$h(\varepsilon,r_{\!*})={{R_{\,\,\varepsilon,s,l}^{(1)+}(r_{\!*})}\big/
{R_{\,\,\varepsilon,s,l}^{(1)-}(r_{\!*})}}$. Then Eq.
(\ref{r*spectra}), written down in the short form
$g(\varepsilon)-h(\varepsilon,r_{\!*})=0$, defines implicitly the
function $\varepsilon(r_{\!*})$ in the general case when
$\partial_\varepsilon
\big(g(\varepsilon)-h(\varepsilon,r_{\!*})\big)\neq 0$. For a
derivative of the function $\varepsilon(r_{\!*})$ one obtains
$\partial_\varepsilon
\big(g(\varepsilon)-h(\varepsilon,r_{\!*})\big){{d\varepsilon}\over{dr}}=\partial
 h(\varepsilon,r_{\!*})=
 -W\left(R_{\,\,\varepsilon,s,l}^{(1)+}(r_{\!*}),R_{\,\,\varepsilon,s,l}^{(1)-}(r_{\!*})\right)
 \big/\left(R_{\,\,\varepsilon,s,l}^{(1)-}(r_{\!*})\right)^2$,
 where
 $W\left(R_{\,\,\varepsilon,s,l}^{(1)+}(r_{\!*}),R_{\,\,\varepsilon,s,l}^{(1)-}(r_{\!*})\right)$
 denotes the corresponding Wronskian. Since
 $\left(R_{\,\,\varepsilon,s,l}^{(1)-}(r_{\!*})\right)^2\neq
 \infty$  for finite $r_{\!*}\in (1,\infty)$,
 we can have ${{d\varepsilon}\over{dr}}(r_{\!*})=0$
 if, and only if
 $W\left(R_{\,\,\varepsilon,s,l}^{(1)+}(r_{\!*}),
 R_{\,\,\varepsilon,s,l}^{(1)-}(r_{\!*})\right)=0$.
 This contradicts to linear independence of the local solutions
 $R_{\,\,\varepsilon,s,l}^{(1)+}(r_{\!*})$ and
 $R_{\,\,\varepsilon,s,l}^{(1)-}(r_{\!*})$ for finite $r_{\!*}\in (1,\infty)$.
$\diamond$\vskip .3truecm

\section{Numerical Calculation of Quasi-Normal Modes of
Static Spherically Symmetric Objects}

\subsection{Transition Coefficients From Limit Procedures}

As we have seen in the previous sections, for determination of QNM
in simple models of compact objects we need to use the transition
coefficients $\Gamma_{Y\pm}^{X\pm}(\varepsilon,s,l)$. Since these
are not known explicitly, one has to apply more sophisticated
methods for solving two-point boundary problems, instead of
employing directly Eqs. (\ref{QNMspectrum}) and (\ref{r*spectra}).

In this section we present a new simple technique for calculation
of quasi-normal modes of Schwarzschild black holes and massive
spherically symmetric bodies, using directly the exact solutions
of the Regge-Wheeler equation, found for the first time in the
present article. We use our own computer code, written in Maple 10
package. There a pioneering implementation in computer algebra
systems of the five Heun functions for complex values of argument
and of corresponding parameters already exist. We use the
numerically well studied case of Schwarzschild black holes to
compare our results with the earlier ones, which were obtained by
making use of two other numerical methods of Leaver and Andersson
\cite{QNMnumeric}.

The basic idea of our new method is to use proper limit procedures
for calculation of transition coefficients.

Let us consider some solution $R_{\varepsilon,s,l}(r)$ of the
stationary Regge-Wheeler equation and let
\ben R_{\varepsilon,s,l}(r)=
C_{1+}R_{\,\,\varepsilon,s,l}^{(1)+}(r)+
C_{1-}R_{\,\,\varepsilon,s,l}^{(1)-}(r)=
C_{\infty+}R_{\,\,\varepsilon,s,l}^{(\infty)+}(r)+
C_{\infty-}R_{\,\,\varepsilon,s,l}^{(\infty)-}(r)\la{Rexpansion}\een
present its expansions on the corresponding basic solutions
(\ref{R1:ab}) and (\ref{Rinf:ab}). Then
\ben
C_{\infty\pm}=C_{1+}\Gamma_{\infty\pm}^{1+}(\varepsilon,s,l)+
C_{1-}\Gamma_{\infty\pm}^{1-}(\varepsilon,s,l),
\la{CC}\een
and we obtain the modulus of these constants using relations
(\ref{limRinf:ab}) in calculation of the following limits:
\begin{subequations}\label{Clim:ab}
\ben |C_{\infty+}|&=&\lim_{|r|\to +\infty}|e^{+i\varepsilon(r+\ln
r)}R_{\varepsilon,s,l}(r)|:\,
\,\,\,\hbox{for}\,\,\,\arg(r)+\arg{\varepsilon}\in (0,\pi),\label{Clim:a}\\
|C_{\infty-}|&=&\lim_{|r|\to +\infty}|e^{-i\varepsilon(r-\ln
r)}R_{\varepsilon,s,l}(r)|:\,
\,\,\,\hbox{for}\,\,\,\arg(r)+\arg{\varepsilon}\in
(-\pi,0).\label{Clim:b}\een
\end{subequations}
In addition, it becomes clear from asymptotics
(\ref{R_asimptotic}) that in the limit (\ref{Clim:b}) we have the
steepest descent, if and only if
$arg(r)+\arg{\varepsilon}=-{\pi\over 2}\,/\!\!\mod 2\pi$. This
direction is the optimal one for numerical calculations of the
above limits, as we shall see below.

This way we can obtain the needed information about transition
coefficients $\Gamma_{\infty\pm}^{\,\,1\,\pm}(\varepsilon,s,l)$,
using any solution with known constants $C_{1\pm}$. Alternatively,
we can impose the corresponding boundary conditions directly on
the modulus $|C_{\infty\pm}|$, calculated via the limits
(\ref{Clim:ab}).

\subsection{QNM of Schwarzschild Black Holes}

Already in a first calculation the QNM frequencies, made by
Chandraseckhar and Detweiler \cite{QNMnumeric} the following
difficulties were recognized: It is hard to nullify numerically
the coefficient in front of the solution
$R_{\,\,\varepsilon,s,l}^{(\infty)-}(r)$ (see Eq.
(\ref{QNMspectrum})), which is small on the left of the infinite
point on the {\em real} axes. This happens because it is hard to
distinguished the small solution
$R_{\,\,\varepsilon,s,l}^{(\infty)-}(r)$ in the presence of
unavoidable numerical errors in the solution
$R_{\,\,\varepsilon,s,l}^{(\infty)+}(r)$, which is dominant in
this domain. Therefore, much more sophisticated methods, like
Leaver's continuous fraction method and Andersson's phase
amplitude method for accurate calculations of QNM spectrum, were
invented successfully, see references in \cite{QNMnumeric}. This
way were obtained the known most accurate values for QNM
frequencies.

For example, in Andersson's article, among many others, one can
find the value of the basic quasi-normal mode of gravitational
waves: $\varepsilon_{1\smile\infty
}(0,2,2)_{Andersson}=0.747343368 +i\,0.17792463$.

Using the Leaver's method on modern computers, it is easy to find
several more figures:
$\varepsilon_{1\smile\infty }(0,2,2)|_{Leaver}=0.747343368836
+i\,0.177924631377$ for the same mode\footnote{The author is
deeply indebted to prof. Kostas Kokkotas for permission to use his
computer code for calculation of this value by Leaver's method.}.

Since today we have at our disposal a possibility to calculate the
canonical solution (\ref{Scan}) of Heun's equation in the complex
domain of variables using Maple 10, we are able to turn back to
the idea of solving numerically Eq. (\ref{QNMspectrum})), in a
properly modified way, which overcomes the difficulty of
Chandraseckhar-Detweiler method. Our solution is based on the
limit procedure (\ref{Clim:b}) which shows that in the direction
$arg(r)+\arg{\varepsilon}=-{\pi\over 2}$ the solutions interchange
their roles. The former small solution becomes a dominant one.
Taking into account expressions (\ref{R1:ab}), one easily obtains
the following form of the equation for for QNM spectrum of
Schwarzschild black holes:
\ben \lim_{|r|\to +\infty}\bigg|\,r^{s+1}
HeunC\left(-2i\varepsilon,2i\varepsilon,2s,-2\varepsilon^2,2\varepsilon^2\!+\!s^2\!-\!l(l\!+\!1),
1\!-\!|r| e^{i\left({\pi \over 2}\!+\!
\arg(\varepsilon)\right)}\right)\!\bigg|\!=\!0.
\la{NumBHspect}\een

In a numerical treatment of the problem we replace the last
limiting procedure with calculation at some  finite point
$r_{\!\infty}\gg 1$, i.e. we use the equation:
\ben \bigg|(r_{\!\infty})^{s+1}
HeunC\left(-2i\varepsilon,2i\varepsilon,2s,-2\varepsilon^2,2\varepsilon^2+s^2-l(l+1),
1-r_{\!\infty} e^{i\left({\pi \over 2}+
\arg(\varepsilon)\right)}\right)\bigg|=0. \la{NumBHspectrum}\een
We apply the complex plot abilities of Maple 10 to find a
relatively small vicinity of the complex roots of this equation.
Then we justify their values to the desired precision using a
direct check of the place of the corresponding root in the complex
plane $\mathbb{C}_\varepsilon$  and/or by Muller's method for
finding of complex roots \cite{Muller}.

The results are shown in the  Tables \ref{table2}, \ref{table3},
and \ref{table4}. In the Table \ref{table2} the results of our
numerical calculations for the first five eigenvalues are
presented. We see an excellent agreement with the most accurate
values, published up to now by Andersson \cite{QNMnumeric}.

\begin{table}[here]~\vspace{.truecm}
\begin{center}
\begin{tabular}{cllllrrrr}
\hline \hline
$n=0$&$\Big|$& 0.747343368&+ i  0.177924631&\\
\hline
$n=1$&$\Big|$& 0.693421994&+ i  0.547829750  &\\
\hline
$n=2$&$\Big|$& 0.602106909&+ i  0.956553966&\\
\hline
$n=3$&$\Big|$& 0.503009924&+ i  1.410296405&\\
\hline
$n=4$&$\Big|$& 0.415029159&+ i  1.893689781&\\
\hline \hline
\end{tabular}
\end{center}
\caption{The first five eigenvalue for 64 digit calculations and
for $r_{\!\infty}=20$.} \la{table2}
\end{table}
\vskip 0.5truecm

The Table \ref{table3} shows that using high precision
calculations with Eq. (\ref{NumBHspectrum}) one can produce the
value of $\varepsilon_{1\smile\infty }(0,2,2)$ up to more figures
-- here up to 12 figures. This consideration inspires confidence
in our new method for calculation of QNM. As seen from Table
\ref{table3},  with our Maple 10 computer code we obtain the right
9 figures already in 64 digits calculations. This gives us more
confidence in the results for next several eigenvalues, obtained
using the same method and shown in Table \ref{table2}.

\begin{table}[here]~\vspace{.truecm}
\begin{center}
\begin{tabular}{cllllrrrr}
\hline \hline
$32$&$\Big|$& 0.747343368602&+ i 0.177924631717&\\
\hline
$64$&$\Big|$& 0.747343368782&+ i 0.177924631969&\\
\hline
$128$&$\Big|$& 0.747343368864&+ i 0.177924631732&\\
\hline
$256$&$\Big|$& 0.747343368836&+ i 0.177924631360&\\
\hline \hline
\end{tabular}
\end{center}
\caption{The basic eigenvalue for higher number of digits (given
in the first column) calculations (no rounding) and for
$r_{\!\infty}=20$.} \la{table3}
\end{table}
\vskip 0.5truecm

In the Table \ref{table4} we show the deviations of the values of
the basic eigenvalue $\varepsilon_{1\smile\infty }(0,2,2)$,
obtained by our method for different $r_{\!\infty}$, with respect
to the corresponding value for $r_{\!\infty}=100$ and for 32
digits calculations.
\vskip 0.5truecm
\begin{table}[here]~\vspace{.truecm}
\begin{center}
\begin{tabular}{cllllrrrr}
\hline \hline
$10$&$\Big|$&-.17e-8 - i .211e-7&$\Big|$ 100& $\Big|$&0 + i 0  &\\
\hline
$20$&$\Big|$&-.90e-8 - i .106e-7&$\Big|$ 200& $\Big|$&.252e-7 - i .141e-7& \\
\hline
$30$&$\Big|$&-.96e-8 - i .105e-7&$\Big|$ 300& $\Big|$&-.156e-7 + i .204e-7& \\
\hline
$40$&$\Big|$&-.65e-8 - i .82e-8 &$\Big|$ 500& $\Big|$&.2007e-6 - i .3180e-6& \\
\hline
$50$&$\Big|$&-.103e-7 - i .160e-7&$\Big|$1000& $\Big|$&-.785e-7 - i .3198e-6& \\
\hline
$80$&$\Big|$&-.94e-8 - i .50e-8& $\Big|$1500&$\Big|$&.2680e-6 - i .21134e-5& \\
\hline \hline
\end{tabular}
\end{center}
\caption{The numerical results for 32 digit calculations of
$\varepsilon_{1\smile\infty }(0,2,2)$ for different values of
$r_{\!\infty}$ (given in the first and in the third column). In
this Table we present only the deviations from the value
$\varepsilon_{1\smile\infty }(0,2,2)|_{100}= 0.747343377577 + i\,
0.177924642293$,} obtained for $r_{\!\infty}=100$.\la{table4}
\end{table}
\vskip 0.5truecm

As we see, the calculation of the eigenvalue
$\varepsilon_{1\smile\infty }(0,2,2)$, based on equation
(\ref{NumBHspectrum}), is  not very sensitive to the choice of
$r_{\!\infty}$ and for $r_{\!\infty}\geq 20$ it gives a stable
right result up to seven figures for 32 digits calculations. A
simple look on the Table \ref{table4} shows that the deviations
have no systematic nature and may demonstrate instabilities (after
the first seven figures) of Maple 10 calculations with 32 digits
precision.

\subsection{QNM of Massive Compact Body}

Applying the limit procedure (\ref{Clim:b}) to our simple model of
massive compact body, described in Section 3.4, we obtain the
following equation for the corresponding QNM:
\ben \Bigg|\lim_{|r|\to +\infty}\Bigg(
e^{2i*\varepsilon\ln\left(\!|r|e^{-i(\pi/2
+\arg(\varepsilon))}\!-\!1\!\right)}\times\hskip 7.9truecm\nonumber\\
{{HeunC\left(\!-2i\varepsilon,+2i\varepsilon,2s,-2\varepsilon^2,2\varepsilon^2\!+\!s^2\!-l(l\!+\!1),1\!-\!
|r|e^{-i(\pi/2\!+\!\arg(\varepsilon))}\!\right)} \over
{HeunC\left(\!-2i\varepsilon,-2i\varepsilon,2s,-2\varepsilon^2,2\varepsilon^2\!+\!s^2\!-\!l(l\!+\!1),1\!-\!
|r|e^{-i(\pi/2\!+\!\arg(\varepsilon))}\!\right)}
}\Bigg)-\\e^{2i*\varepsilon\ln\left(\!r_{\!*}\!-\!1\!\right)}\times
{{HeunC\left(\!-2i\varepsilon,+2i\varepsilon,2s,-2\varepsilon^2,2\varepsilon^2\!+\!s^2\!-l(l\!+\!1),1\!-\!
r_{\!*}\!\right)} \over
{HeunC\left(\!-2i\varepsilon,-2i\varepsilon,2s,-2\varepsilon^2,2\varepsilon^2\!+\!s^2\!-\!l(l\!+\!1),1\!-\!
r_{\!*}\!\right)}}\Bigg|=0.\hskip .5truecm\nonumber
 \la{NumPPspect}\een

\begin{figure}[htbp] \vspace{6.truecm}
\includegraphics{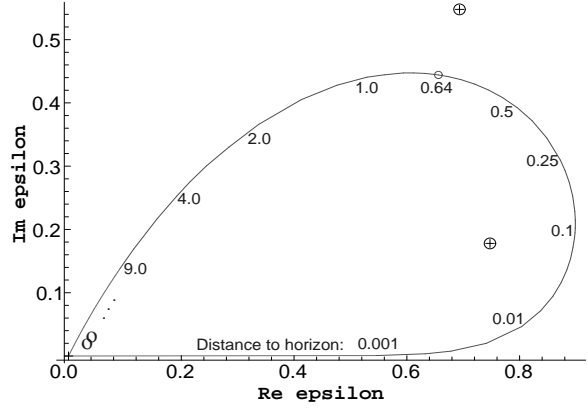} \caption{\hskip 0.2truecm The complex trajectory of the
basic eigenvalue $\varepsilon_{ r_{\!*}\,\bullet\hskip
-0.1truecm\thicksim\, \infty }(0,2,2)$  of QNM of compact massive
body as a function of the distance $d=r_{\!*}-1>0$ of the body's
surface to the horizon (in units $2M$). For comparison purposes,
the first two eigenvalues of QNM for Schwarzschild black holes
with the same mass $M$ are shown, using the circles with crosses.
The origin of the complex plane $\mathbb{C}_\varepsilon$ is marked
by a small cross. It presents simultaneously the event horizon
(where the trajectory begins) and the physical infinity (where the
trajectory ends). The small circle on the trajectory denotes the
position of the maximum of the effective potential $V(r)$.
    \hskip 1truecm}
    \label{Fig2}
\end{figure}

\begin{figure}[htbp] \vspace{6.truecm}
\includegraphics{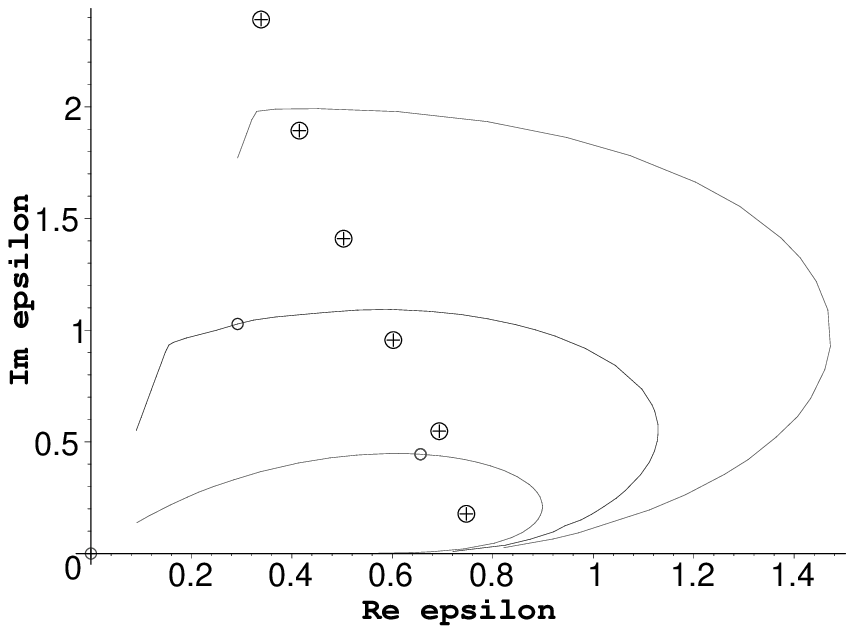} \caption{\hskip 0.2truecm The trajectories in the
complex plane $\mathbb{C}_\varepsilon$ of the first three
eigenvalues: $\varepsilon_{ r_{\!*}\,\bullet\hskip
-0.1truecm\thicksim\, \infty }(0,2,2)$,$\varepsilon_{
r_{\!*}\,\bullet\hskip -0.1truecm\thicksim\, \infty }(1,2,2)$ and
$\varepsilon_{ r_{\!*}\,\bullet\hskip -0.1truecm\thicksim\, \infty
}(2,2,2)$ of QNM of massive compact body. For comparison purposes,
the first six eigenvalues of QNM for Schwarzschild black holes
with the same mass $M$ are shown, using the circles with crosses.
The small circles on the trajectories denote the corresponding
positions of the maximum of effective potential $V(r)$.
    \hskip 1truecm}
    \label{Fig3}
\end{figure}

The numerical treatment of this problem is the same as in the
previous Subsection: we once more replace the limiting procedure
with calculation at some finite point $r_{\!\infty}\gg 1$, i.e. we
use the equation:
\ben \Bigg|e^{2i*\varepsilon\ln\left(\!r_{\!\infty}e^{-i(\pi/2
+\arg(\varepsilon))}\!-\!1\!\right)}\times\hskip 9.2truecm\nonumber\\
{{HeunC\left(\!-2i\varepsilon,+2i\varepsilon,2s,-2\varepsilon^2,2\varepsilon^2\!+\!s^2\!-l(l\!+\!1),1\!-\!
r_{\!\infty}e^{-i(\pi/2\!+\!\arg(\varepsilon))}\!\right)} \over
{HeunC\left(\!-2i\varepsilon,-2i\varepsilon,2s,-2\varepsilon^2,2\varepsilon^2\!+\!s^2\!-\!l(l\!+\!1),1\!-\!
r_{\!\infty}e^{-i(\pi/2\!+\!\arg(\varepsilon))}\!\right)} }-\hskip
1.3truecm\\e^{2i*\varepsilon\ln\left(\!r_{\!*}\!-\!1\!\right)}\times
{{HeunC\left(\!-2i\varepsilon,+2i\varepsilon,2s,-2\varepsilon^2,2\varepsilon^2\!+\!s^2\!-l(l\!+\!1),1\!-\!
r_{\!*}\!\right)} \over
{HeunC\left(\!-2i\varepsilon,-2i\varepsilon,2s,-2\varepsilon^2,2\varepsilon^2\!+\!s^2\!-\!l(l\!+\!1),1\!-\!
r_{\!*}\!\right)}}\Bigg|=0\nonumber \la{NumPPspectrum}\een
and apply our own computer codes, written in Maple 10, for finding
its roots. The results are shown in  Fig. \ref{Fig2} and Fig.
\ref{Fig3}.

\section{Conclusion}

In the present article we have solved some mathematical problems
related to exact solution of the Regge-Wheeler equation. Since the
obtained results and possible further developments were described
in detail in the basic text, in the concluding remarks we will
comment on their eventual physical significance and some open
problems.

In the present article we do not establish a definite relation
between the spectrum of the problem with Dirichlet's boundary
condition at the surface of the spherically symmetric body and the
corresponding spectrum of more realistic models of relativistic
stars \cite{QNMinStars}. The simple comparison of the published
numerical results for relativistic stars with our results in Figs.
\ref{Fig2}, \ref{Fig3} shows that the complex eigenvalues of
star's axial modes are placed in a complicated way in the domains,
separated by the trajectories of QNM eigenvalues $\varepsilon_{
r_{\!*}\,\bullet\hskip -0.1truecm\thicksim\, \infty }(n,s,l)$,
which belong to the simple model of massive compact objects,
studied here for the first time\footnote{The author is grateful to
unknown referee for its suggestion to consider such comparison.}.
One may hope that further investigation of this problem will help
our understanding of QNM of compact matter objects.

A more realistic model for real spherically symmetric bodies may
be obtained considering as a boundary condition a partial
reflection and partial penetration of the waves through the body's
surface. In the present article we have considered only the two
extreme cases for reflection coefficient: $R=0$ -- for the black
holes and $R=1$ -- for our simple model of bodies. It is
interesting to know will it be possible to connect continuously
these two cases, introducing a coefficient of reflection $R\in
[0,1]$, or the difference in the boundary conditions is an
insuperable obstacle for this.

In the near future our results may be useful for the planned
observations of gravitational waves emitted by or spreading around
existing real compact objects. These results may help to clarify
the physical nature of the observed very heavy and very compact
dark objects in the universe. Our consideration gives a unique
possibility for {\em a direct} experimental test of the existence
of space-time {\em holes}. The study of spectra of the
corresponding waves, propagating around compact dark objects, may
give indisputable evidences, whether there is space-time hole
inside such an invisible object.

Similar results can be obtained for electromagnetic, spinor and
scalar waves in the external Schwarzschild gravitational field of
compact objects. A further study of these phenomena is an
important physical problem and may offer practical physical
criteria for experimental distinction of the two completely
different hypothetical models:

1) The model of space-time black holes; and

2) Different new models of very compact dark objects, made of {\em
real}, or of some {\em hypothetical} matter \cite{Fiziev}.

The present-day astrophysical observations still are not able to
make difference between these two cases. The only real
observational fact is that we see very compact and very massive
objects which show up {\em only} due to their {\em external}
strong gravitational field. An actual theoretical problem is to
find a convincing model for description of these already observed
compact dark objects and the criteria for experimental
verification of a model like this. We hope that the present
article is a step in this direction.
\vskip 1truecm

{\em \bf Acknowledgments} \vskip .3truecm

The author is grateful to the High Energy Physics Division, ICTP,
Trieste, for the hospitality and for the nice working conditions
during his visit in the autumn of 2003. There the idea of the
present article was created. The author is grateful to the JINR,
Dubna, too, for the priority financial support of the present
article, for the hospitality and for the good working conditions
during his three months visits in 2003, 2004 and 2005, when the
study was performed in detail. This article also was supported by
the Scientific Found of Sofia University, and by its Foundation
"Theoretical and Computational Physics and Astrophysics".

I wish to acknowledge the participants in the scientific seminars
of ICTP -Trieste, BLTF of JINR - Dubna and INRNE - Sofia for
stimulating discussions of the basic ideas, methods, and results
of the present article.

I am deeply indebted to prof. Kostas Kokkotas for the useful
discussions and for providing a simple FORTRAN computer code for
calculations of the first several QNM of Schwarzschild black holes
by Leaver's method. This code, being produced and used in earlier
studies in his group in Thessaloniky  was kindly e-mailed by prof.
Kokkotas to our Joint Group on Gravity and Astrophysics in Sofia.
The comparison of our Maple 10 results for Schwarzschild black
holes with the corresponding high precision results, obtained by
the Leaver's method, increases substantially our confidence in the
new method, developed in the present article.

I wish to express my thanks to prof. Michail Todorov -- for help
and discussions of numerical techniques used in the preset
article, to prof. Nils Andersson -- for sending me a copy of his
article, cited in \cite{QNMnumeric}, to prof. Matt Visser for
drawing my attention to five more references, included in
\cite{Fiziev}, to prof. Pawel Mazur -- for adding a new reference
and corrections in \cite{Fiziev}, and to two unknown referees for
their extremely useful suggestions for improvements of the article
and for help in the references.

\section{ Appendix: General Properties of the Complex
Current and Some Important Consequences}

\subsection{General Properties of the Complex
Curl}

At first glance the Regge-Wheeler eigenvalue problem seems to be
quite similar to the Schr\"odinger eigenvalue problem in quantum
mechanics. There one considers usually {\em real} eigenvalues of
the energy operator. The last assumption yields existence of a
conserved {\em real} current. This current describes an essential
physical information about the solutions of the problem. We were
not able to find in the literature a proper investigation of the
properties of the current in RW problem with {\em complex}
eigenvalues $\varepsilon$. Therefore in this Appendix we define
the corresponding {\em complex} current, study its basic
properties and derive some consequences, which are essential for
our article.

In the case of $\varepsilon\in \mathbb{C}$, $\varepsilon_R>0$,
$\varepsilon_I\neq 0$, and $V\in \mathbb{R}$ the stationary RW
equation (\ref{R}) for $R=X+iY=|R|e^{i\varphi}$ is equivalent to
the real system of 4th order ordinary differential equations:
\ben X^{\prime\prime}=(V-\xi)X+\eta
Y,\,\,\,Y^{\prime\prime}=(V-\xi)Y-\eta X\la{XY},\een
where the prime denotes differentiation with respect to RW
tortoise variable $x$ and
$\varepsilon^2=\varepsilon_R^2-\varepsilon_I^2+2i\varepsilon_R\varepsilon_Y=\xi+i\eta$.
In the case $\eta\neq 0$ this system does not split into two
independent differential equations of 2nd order and in addition we
obtain:
\ben \xi = V+{{(X^\prime)^2+(Y^\prime)^2}\over{X^2+Y^2}}-
{{\left(XX^\prime+YY^\prime\right)^\prime}\over{X^2+Y^2}},
\,\,\,\eta={{\left(YX^\prime-XY^\prime\right)^\prime}\over{X^2+Y^2}}.\la{xieta}\een

Despite of the above fact, for $\eta\neq 0$ the system (\ref{XY})
is equivalent to 2nd order system of {\em real} nonlinear
differential equations (see, for example, the article by
S.~Chandrasekhar and S.~L.~Detweiler in \cite{QNMnumeric}):
\ben u_R^\prime=\eta-2u_Ru_I,\,\,\,u_I^\prime=u_R^2-u_I^2+V-\xi,
\la{uRuI}\een
with  boundary conditions at infinity:
$u_R(\infty)=\varepsilon_R>0,\,\,\,u_I(\infty)=\varepsilon_I>0$ --
for QNM. This phenomenon turns to be possible since two of the
arbitrary real constants in the general solution of the system
(\ref{XY}) are actually not essential, entering inessential
complex constant multiplier of form $R_*=|R_*|e^{i\varphi_*}$ in
the solution $R$ of the RW eq. (\ref{R}). This constant always can
be chosen to be equal to 1. To obtain Eqs. (\ref{uRuI}) we use the
substitution:
\ben R=e^{-i\int u(x) dx}= e^{-i\int u_R(x) dx}e^{\int u_I(x)dx},
\la{Ru}\een
where
\ben u=u_R+iu_I=i{{R^\prime}\over{R}}=i{{R^\prime}\bar
R\over{|R|^2}}={{j_R+i\, j_I}\over{|R|^2}}.\la{uj}\een

This way we introduce a {\em complex} current $j=j_R+i
\,j_I=u_R|R|^2+i\, u_I|R|^2$ in the RW problem. Using the Eqs.
(\ref{XY}), it is easy to obtain the following relations:
\ben j_R=YX^\prime-XY^\prime,\,\,\,j_R^\prime=\eta(X^2+Y^2),\,\,\,
j_R^{\prime\prime}=2\eta(XX^\prime+YY^\prime),\nonumber\\
j_R^{\prime\prime\prime}=
2\eta\left((X^\prime)^2+(Y^\prime)^2\right)+2\eta(V-\xi)(X^2+Y^2),\hskip
1.5truecm  \la{djR}\een\
and the 4th order differential equation for the real part $j_R$ of
the complex current
\ben j_R^{IV}=4(V-\xi)j_R^{\prime\prime}+V^\prime
j_R^{\prime}+4\eta^2j_R.\la{4jeq}\een
One sees that for $\eta\neq 0$:

i) The complex current $j$ is not conserved, i.e. $j_R\neq const$
and $j_I\neq const$.

ii) Since $j_I={1\over
{2\eta}}j_R^{\prime\prime}=XX^\prime+YY^\prime$, in the imaginary
part $j_I$ of of the current $j$ there is no independent
information.

iii) In addition $|R|^2=j_R^\prime/\eta$, $\varphi^\prime=-\eta
j_R/j_R^\prime$. Hence, the real component $j_R$ of the complex
current $j$ contains all essential information for the RW problem
and the Eq. (\ref{4jeq}) can replace RW one (\ref{R}) in all
considerations.

It is easy to obtain the first integral
${1\over 4}\left((j_R^\prime)^2\right)^{\prime\prime}-{3\over
4}(j_R^{\prime\prime})^2-(V-\xi)(j_R^{\prime})^2-\eta^2 j_R^2
=const$ of the Eq. (\ref{4jeq}). The constant value of this
integral turns to be one of the two inessential constants in the
general solution of the Eq. (\ref{4jeq}). Without loss of
generality we can assume that this constant equals zero. Then the
real part $j_R$ obeys 3th order differential equation:
\ben {1\over 4}\left((j_R^\prime)^2\right)^{\prime\prime}={3\over
4}(j_R^{\prime\prime})^2+(V-\xi)(j_R^{\prime})^2+\eta^2j_R^2\la{3jeq}\een
Using the substitution $j_R=exp\left(\eta\int u_R(x)dx\right)$ one
can exclude the second inessential constant in the general
solution of the Eq. (\ref{4jeq}), lowering once more its order.
Thus we obtain the single 2nd order differential equation:
\ben {1\over 2} \left({{\eta-u_R^\prime}\over{u_R}}\right)^\prime=
-{1\over 4}
\left({{\eta-u_R^\prime}\over{u_R}}\right)^2+u_R^2+V-\xi,
\la{2jR}\een
which can be derived directly from the system (\ref{uRuI}), too.
The last observation supports our choice of the zero value for the
first integral of Eq. (\ref{4jeq}).

At the end let us introduce functions
$R_{\leftarrow}={1\over 2}\left(R+{1\over{ik}}R^\prime\right)$,
$R_{\rightarrow}={1\over 2}\left(R-{1\over{ik}}R^\prime\right)$
putting
\ben R=R_{\leftarrow}+R_{\rightarrow},\,\,\, R^\prime=ik\left(
R_{\leftarrow}-R_{\rightarrow}\right).\la{RlRr}\een
Here $k=\sqrt{\varepsilon^2-V}$ and we suppose to choose the
branch of the square root for which we have $k=\varepsilon$  when
$V=0$. Now we obtain
\ben j=
k\left(|R_{\rightarrow}|^2-|R_{\leftarrow}|^2\right),\la{jRlRr}\een
precisely as in the real case $\varepsilon_I=0$.

The complex functions $R_{\rightarrow}$ and $R_{\leftarrow}$ obey
the 2nd order system
\ben \left( \begin{array}{cc} R_{\leftarrow}\\ R_{\rightarrow}
\end{array} \right)^\prime= \hat H \left( \begin{array}{cc} R_{\leftarrow}\\ R_{\rightarrow}
\end{array} \right),\,\,\,\,\,\hat H=k\left( \begin{array}{cc}
1&0\\0&1
\end{array} \right)+i {{k^\prime}\over{2 k}}\left( \begin{array}{cc}
1&-1\\-1&1
\end{array} \right),\la{RlRreq}\een
which shows that at the points where $V^\prime=0$ these functions
describe waves which run to the right and to the left,
correspondingly. If, in addition, $V=0$, then
$j_{\rightarrow}=\varepsilon_R|R_{\rightarrow}|^2$,
$j_{\leftarrow}=-\varepsilon_R|R_{\leftarrow}|^2$.

\subsection{Special Properties of the Current in the Case of Compact Body with Dirichlet's
Boundary Condition at the Surface}

In the simple model of compact body with Dirichlet's boundary
condition (\ref{Dirichlet}) one has $R(x_*)=0$,
$R^\prime_*=R^\prime(x_*)\neq 0$. As a result in vicinity of the
point $x_*$ one obtains the following series expansions with
respect to the difference $\Delta x=x-x_*$:
\ben R(x)=R^\prime_*\left(\Delta x +{1\over
6}(V_*-\varepsilon^2)\Delta x^3+ {1\over 12}V^\prime_*\Delta
x^4+{\cal O}_6(\Delta x) \right),\la{serR}\een
\begin{subequations}\label{ser_u:ab}
\ben u_R(x)={\eta\over 3}\Delta x+{\cal O}_3(\Delta x)
,\hskip 4.6truecm\label{ser_u:a}\\
u_I(x)={1\over \Delta x} +{1\over 3}(V_*-\xi)\Delta x+ {1\over
4}V^\prime_*\,\Delta x^2+{\cal O}_3(\Delta x); \label{ser_u:b}\een
\end{subequations}
\begin{subequations}\label{ser_j:ab}
\ben j_R(x)=\eta|R^\prime_*|^2\left({1\over 3}\Delta x^3 +{1\over
15}(V_*-\xi)\Delta x^5+ {1\over 36}V^\prime_*\Delta
x^6+{\cal O}_7(\Delta x) \right),\label{ser_j:a} \\
j_I(x)=|R^\prime_*|^2\left(\Delta x +{2\over 3}(V_*-\xi)\Delta
x^3+ {5\over 6}V^\prime_*\Delta x^4+{\cal O}_5(\Delta x) \right);
\hskip 1.1truecm\label{ser_j:b}\een \end{subequations}
and the two equivalent sets of integral representations:
\begin{subequations}\label{int_uRuI:ab}
\ben
u_R(x)=\eta|R(x)|^{-2}\int\limits^x_{x_*}|R(x)|^2dx,\hskip 5.7truecm\label{int_uRuI:a}\\
u_I(x)=|R(x)|^{-1}\int\limits^x_{x_*}\left(|u_R(x)|^2+\big(V(x)-\xi\big)|R(x)|\right)dx+
{{|R^\prime_*|}\over{|R(x)|}}, \label{int_rRuI:b}\een
\end{subequations}

and
\begin{subequations}\label{int_jRjI:ab}
\ben
j_R(x)=\eta\int\limits^x_{x_*}|R(x)|^2dx,\hskip 5.9truecm\label{int_jRjI:a}\\
j_I(x)=\int\limits^x_{x_*}\left(|R^\prime(x)|^2+V(x)|R(x)|^2\right)dx-\xi\int\limits^x_{x_*}|R(x)|^2dx.
\label{int_jRjI:b}\een \end{subequations}

With the help of the Eq. (\ref{int_jRjI:a}) the Eq.
(\ref{int_jRjI:b}) can be rewritten in the form
\ben \xi j_R(x)+\eta j_I(x)=\eta
\int\limits^x_{x_*}\left(|R^\prime(x)|^2+V(x)|R(x)|^2\right)dx.
\la{jRjI_rel}\een

Now we obtain the following additional properties of the solutions
of RW equation for compact body with Dirichlet's boundary
condition at its surface:

1. NM do not exist for $\eta\neq 0$ in the Dirichlet's boundary
problem (\ref{Dirichlet})  . Indeed, since for NM $R(\infty)=0$,
one must have for them
$0=j_R(\infty)=\eta\int\limits^\infty_{x_*}|R(x)|^2dx$. Hence,
$R(x)\equiv 0$ for $x\in [x_*,\infty)$ and we will have a trivial
zero-solution.

2. For QNM we have $\eta>0$, because of our assumption
$\varepsilon_R>0$. Then we obtain the following properties of QNM
in this problem:

i) The solutions $R(x)$ and the current $j_R(x)$ cannot have other
zeros $x_0$, different from $x_*$. Indeed, if $R(x_0)$=0, then
$0=j_R(x_0)=\eta\int\limits^{x_0}_{x_*}|R(x)|^2dx\geq 0$ leads to
$x_0\equiv x_*$ for any nontrivial solution $R(x)\neq 0$.

ii) If $\varepsilon_I\geq \varepsilon_R$, then $\xi\leq 0$ and Eq.
(\ref{int_jRjI:b}) shows that $j_I(x)>0$ and $j_r^{\prime\prime
}>0$ for any $x\in (x_*,\infty)$.

iii) Oscillations of $j_I(x)$ (and $j_r^{\prime\prime }$) around
the zero value:

From Eq. (\ref{int_jRjI:b}) we see, that $j_I(x)$ (and
$j_R^{\prime\prime}(x)=2\eta j_I(x)$) can have a second zero,
besides $x_*$, if and only if $\varepsilon_I < \varepsilon_R$,
i.e. when $\xi>0$. Then, since $\lim\limits_{x\to\infty}j_R(x)=
\lim\limits_{x\to\infty}\left(u_R(x)|R(x)|^2\right)=+\infty$, the
current $j_R(x)$ must have at list one more (third) zero in the
interval $[x_*, \infty)$. This way we see that in general case for
$\xi>0$ the imaginary part $j_I(x)$ of the complex current has an
odd number ($\geq 1$) of zeros and its derivative $j_I^\prime(x)$
has an even number ($\geq 0$) of zeros in the interval $[x_*,
\infty)$.

iv) From Eq. (\ref{jRjI_rel}) for $x>x_*$ we obtain the
inequality:
\ben j_I(x)\geq
{{1}\over{2}}\left({{\varepsilon_I}\over{\varepsilon_R}}-
{{\varepsilon_R}\over{\varepsilon_I}}\right)j_R(x)\la{jIjR_ineq}.\een

\end{document}